\documentclass[aps,prl,reprint,superscriptaddress,letterpaper,twocolumn,longbibliography]{revtex4-2}
\usepackage[english]{babel}
\usepackage{ucs}
\usepackage[utf8x]{inputenc}
\usepackage{amsmath}
\usepackage{amssymb}
\usepackage{amsthm}
\usepackage{amsfonts}
\usepackage{mathtools}
\usepackage{mathrsfs}
\usepackage{graphicx}
\usepackage{bm}
\usepackage{bbm}
\usepackage{empheq}
\usepackage{cases}
\usepackage{euscript}
\usepackage[usenames, dvipsnames, x11names]{xcolor}
\usepackage[colorlinks=true,linkcolor=SpringGreen4,citecolor=blue,urlcolor=Magenta3]{hyperref}
\usepackage[shortlabels]{enumitem}
\usepackage[draft,todonotes={textsize=scriptsize}]{changes}
\definechangesauthor[name={Eva Weig}, color=violet]{EW}
\definechangesauthor[name={Hugo Ribeiro}, color=SpringGreen4]{HR}
\definechangesauthor[name={Avishek Chowdhury}, color=Blue4]{AC}
\definechangesauthor[name={Anh-Tuan Le}, color=RedOrange]{ATL}
\newcommand{\mm}[1]{\mathrm{#1}}
\newcommand{\abs}[1]{\left|#1\right|}

\newcommand{\di}[1]{\mathop{}\!\mathrm{d} #1}

\newcommand{\avg}[1]{\langle #1 \rangle}

\newcommand{\ceil}[1]{\lceil {#1} \rceil}
\newcommand{\floor}[1]{\lfloor {#1} \rfloor}
\def \uf{\mathrm{f}}

\def \ur{\mathrm{r}}
\def \us{\mathrm{s}}

\def \uI{\mathrm{I}}

\def \uw{\mathrm{w}}

\def \av{\mbox{\boldmath$a$}}

\def \nv{\mbox{\boldmath$n$}}
\def \sigmav{\mbox{\boldmath$\sigma$}}

\def \nvs{\mbox{\boldmath$\scriptstyle{n}$}}

\def \sigmav{\mbox{\boldmath$\sigma$}}
\DeclareFontFamily{OT1}{pzc}{}
\DeclareFontShape{OT1}{pzc}{m}{it}{<-> s * [1.10] pzcmi7t}{}
\DeclareMathAlphabet{\mathpzc}{OT1}{pzc}{m}{it}

\begin{document}

\title{Iterative Adaptive Spectroscopy of Short Signals}
\author{Avishek Chowdhury}
\affiliation{ 
	Department of Electrical \& Computer Engineering, Technical University of Munich, 80333 Munich, Germany}
	
\author{Anh Tuan Le}
\affiliation{ 
	Department of Electrical \& Computer Engineering, Technical University of Munich, 80333 Munich, Germany}
	
\author{Eva M. Weig}
\affiliation{ 
	Department of Electrical \& Computer Engineering, Technical University of Munich, 80333 Munich, Germany}
\affiliation{
    Munich Center for Quantum Science and Technology (MCQST), Schellingstr. 4, 80799 Munich, Germany}

\author{Hugo Ribeiro}
\affiliation{Department of Physics and Applied Physics, University of Massachusetts Lowell, Lowell, MA 01854, USA}

\begin{abstract}
	We develop an iterative, adaptive frequency sensing protocol based on Ramsey interferometry of a two-level system. Our scheme allows
	one to estimate unknown frequencies with a high precision from short, finite signals. It avoids several issues
	related to processing of decaying signals and reduces the experimental overhead related to sampling. High precision is achieved
	by enhancing the Ramsey sequence to prepare with high fidelity both the sensing and readout state and by using an iterative
	procedure built to mitigate systematic errors when estimating frequencies from Fourier transforms.
\end{abstract}

\maketitle

\textit{Introduction ---}
The precision of any coherent sensor, classical or quantum, is limited by the maximal measurement time-window $t_\uw$ over which a
signal of interest can be sampled. This limitation originates from unwanted interactions of the sensor with noisy environmental
degrees of freedom, which lead to decoherence and thus, to the decay of the measured signal
\footnote{Note that the measured signal (or signal) refers to the data recorded by the sensor, whereas the signal of interest
represents the quantity to be detected.}.

Strategies extending the coherence time~\cite{Fisk1995,Saeedi2013,Zhong2015,Barfuss2018,Kobl2019,Herbschleb2019} or using states
with no classical analogs~\cite{Tse2019,Malnou2019,Abadie2011,Aasi2013,Werninghaus2021,Lawrie2019}, e.g., entangled or squeezed
states, to improve the precision have been put forward. However, these strategies do not guarantee that the sensor operates with
the highest precision attainable. This is only possible if all steps involved in the sensing protocol are
optimized~\cite{liu2022}.

\begin{figure}
	\includegraphics[width=\columnwidth]{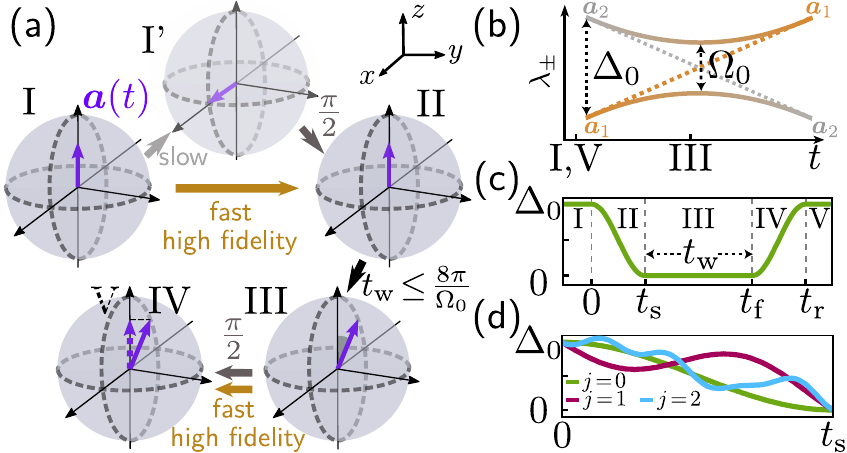}
	\caption{Enhanced Ramsey sequence. (a) The five steps of a Ramsey sequence~\cite{Degen2017} to sense the
		coupling strength $\Omega_0/2$ between a two-mode (or two-level) system as depicted in (b). The normal (slow)
		sequence is indicated by gray arrows, while the enhanced (fast, high-fidelity) sequence follows the orange arrows.
		(c) Typical time-dependent frequency-sweep to generate a Ramsey sequence. (d) Comparison of the leading edge of
		the uncorrected
		(green) and modified (red and blue) frequency-sweeps allowing one to prepare the sensing state [step
		(II) in (a)].
	}
	\label{fig:1} 
\end{figure}

In this work, we demonstrate how to enhance the ubiquitous Ramsey interferometry of a two-level system~\cite{Degen2017} (see
Fig.~\ref{fig:1}) to estimate an unknown frequency from a signal with a short $t_\uw$. There are two reasons to impose a short
$t_\uw$:~First, it is experimentally simpler to sample short signals (smaller number of datapoints required) and secondly, from a
signal processing point of view, it is easier to deal with non-decaying signals to enhance the frequency estimation through the
application of window functions~\cite{Prabhu2014}
\footnote{The decay envelope of the signal can be viewed as an ``uncontrolled'' window function applied to the signal oscillating
component. The presence of an ``uncontrolled'' window function can negate all the benefits of using specific window functions
tailored to enhance frequency estimation.}.

Enhancing Ramsey interferometry is done by reducing the error in preparing both the sensing and readout states [steps (II) and
(IV) in Fig.~\ref{fig:1}~(a)]. To find frequency-sweeps yielding high fidelity state preparation, we use the recently proposed
Magnus-based strategy for control~\cite{Ribeiro2017,Roque2021}.

To improve frequency estimation, one needs to overcome two main shortcomings associated to discrete Fourier transforms:~(FT1)
Spectral leakage~\cite{Harris1978}, which makes a simple, single frequency spectrum appear with multiple frequency components and
can lead to a shift of the maxima of the Fourier transform, and (FT2) finite frequency
resolution~\cite{whittaker1915,nyquist1928,shannon1949}, which follows from having signals that are a discrete, finite set of
points separated by a finite time-interval. We stress that (FT1) is not an issue when $1/t_\uw$ is much smaller than the frequency
we seek to estimate. The Fourier uncertainty principle~\cite{Hogan2005} guarantees then that the resulting signal spectrum in
frequency domain is highly localized.

To mitigate systematic errors originating from (FT1) and (FT2) we develop an iterative, adaptive sensing protocol.  Each
interaction consists in performing Ramsey interferometry and processing the acquired signal with a window function tailored to
either enhance the amplitude of the Fourier spectrum or frequency estimation. After each iteration, we use the newly found
frequency estimate to update steps II, III, and IV of the Ramsey sequence [see Fig.~\ref{fig:1}~(a) and (c)].

Updating steps II and IV entails running the Magnus-based strategy for control with the new frequency estimate to enhance the
frequency-sweeps used to prepare the sensing and readout state. Updating step III requires one to adjust $t_\uw$ to reduce the
effects of spectral leakage. After just a few iterations we reach the spectral resolution allowed by the chosen, short $t_\uw$ and
obtain a high precision frequency estimate.

\textit{Model ---}
We consider a parametrically coupled two-mode system as the coherent sensing element [see Fig.~\ref{fig:1}~(b)].  The dynamical
matrix (effective Hamiltonian) describing the coupled modes is given by (see, e.g., Ref.~\cite{Seitner2017b})
\begin{equation}
	D (t) =\frac{1}{2}\left(\Delta(t)\sigma_z + \left[\Omega_0 + \delta \Omega (t) \right] \sigma_x\right),
	\label{eq:DynMat}
\end{equation}
where $\sigma_j$, $j \in \{x,y,z\}$, are Pauli matrices, $\Delta (t) = \omega_2 (t) -\omega_1 (t)$ is the controllable frequency
difference between mode 1 and 2 [see Fig.~\ref{fig:1}~(b)], $\Omega_0/2$ is the unknown coupling strength we seek to estimate [see
Fig.~\ref{fig:1}~(b)], and $\delta \Omega (t)$ is a real stochastic process describing how noise arising, e.g., from thermal
fluctuations or fluctuations of the fields the system is subjected to, affects the coupling between the modes. Neglecting noise,
the eigenfrequencies of Eq.~\eqref{eq:DynMat} are given by $\lambda_\pm (t) = \pm \sqrt{\Delta^2 (t) + \Omega_0^2}/2$ [see
Fig.~\ref{fig:1}~(b)].

The dynamical matrix $D(t)$ is formally equivalent to a two-level Hamiltonian
$\hat{H}(t)$~\cite{Novotny2010,Faust2013,Frimmer2014a} subject to classical noise. The model we use, thus, describes both coherent
classical
systems~\cite{Seitner2017b,Faust2012b,Okamoto2013a,Braakman2018,Teufel2011,Ranfagni2021,Yeo2014,Thon2009,Perpeintner2018} and
quantum systems~\cite{Bennett2001,koch2007,Rabl2010,Barson2017,Kurpiers2018,Zrenner2002,Boss2017,Poggiali2018}.

Since our goal is to develop a sensing protocol whose duration $t_\ur$ [see Fig.~\ref{fig:1}~(c)] is short, i.e., $(\Omega_0/2\pi)
t_\ur < 5$, we assume that $\delta \Omega (t) \approx \delta \Omega$ does not change appreciably for one realization of the
protocol (frozen environment approximation). Within this framework, averaged Ramsey signals are obtained by performing statistical
averaging, i.e., 
\begin{equation}
	\avg{s (t)}_{\delta \Omega} = \int_{-\infty}^{\infty} \di{\delta\Omega} p(\delta\Omega) s(t),
	\label{eq:NoiseAvg}
\end{equation}
where $p(\delta\Omega)$ is the probability distribution of $\delta\Omega$. Here, we assume $p(\delta\Omega)$ to be a Gaussian
distribution with zero mean and standard deviation $\sigma_{\delta\Omega}$,
\begin{equation}
	p(\delta\Omega)=\frac{1}{\sqrt{2\pi}
	\sigma_{\delta\Omega}}\exp\left[-\frac{1}{2}\frac{\delta\Omega^{2}}{\sigma_{\delta\Omega}^2}\right].
	\label{eq:9}
\end{equation}
In this work we choose $\sigma_{\delta\Omega} =\Omega_0/10$.

\textit{Sensing and readout state preparation ---}
We consider a generic frequency-sweep for the Ramsey interferometry given by [see Fig.~\ref{fig:1}~(c)]
\begin{equation}
	\Delta (t) =
	\begin{cases}
		\Delta_\us (t) = \Delta_0 f (t), &\text{for } 0\leq t \leq t_\us, \\
		0, &\text{for } t_\us < t < t_\uf, \\
		\Delta_\ur (t) = \Delta_0 [1-f (t - t_\uf)], &\text{for } t_\uf \leq t \leq t_\ur,\\
	\end{cases}
	\label{eq:GenDetRamsey}
\end{equation}
where $\Delta (0) = \Delta(t_\ur) = \Delta_0$ is the initial (final) value of the frequency difference [see Fig.~\ref{fig:1}~(b)]
and $f(t)$ is a smooth sweep function obeying $f(0)=f(t_\ur)=1$ and $f(t_\us) = f(t_\uf)=0$. We define the measurement time-window
$t_\uw = t_\uf - t_\us$ and the total sensing time $t_\ur = 2 t_\us + t_\uw$, where we choose the duration of both the sensing and
readout state preparation protocols to be $t_\us$. [see Fig.~\ref{fig:1}~(c)]

By choosing $\Delta_0 \gg \Omega_0$, we can initialize the system in mode $\av_1 = (0,1)^\mathsf{T}$ at $t=0$ [see step (I) in
Figs.~\ref{fig:1}~(a) and (c)]. This is also the sensing state $\av_\us$ we would like to use to probe $\Omega_0$, i.e., $\av_\us
= \av_1$. At the avoided crossing, $\av_1$ can be expressed as an equal superposition of the eigenmodes of Eq.~\eqref{eq:DynMat},
which is the state maximizing the visibility of the Ramsey fringes (see Supplemental Material).

To prepare the sensing state $\av_\us = \av_1$ at $t=t_\us$ and the readout state $\av_\ur$ at $t=t_\ur$ (step II and IV in
Figs.~\ref{fig:1}~(a) and (c), respectively) one needs to choose $f(t)$ such that the evolution generated by Eq.~\eqref{eq:DynMat}
corresponds to the identity in the intervals $0\leq t \leq t_\us$ and $t_\uf \leq t \leq t_\ur$. This can theoretically be
realized with a frequency-sweep whose leading and trailing edges duration fulfills the condition $\Omega_0 t_\us \ll 1$ (quasi
instantaneous sweep). However, faithful reproduction of fast sweeps in the lab environment are limited by the maximum bandwidth of
wave generators. The Landau-Zener model~\cite{landau1932,zener1932,stuckelberg1932,majorana1932}, where one sets the sweep
function to be linear in time, $f (t) = 1 - t/t_\us$, is a perfect example:~Faster and faster sweeps require more and more Fourier
components to accurately reproduce $f (t)$ (see Supplemental Material).

The alternative is to use a frequency-sweep with an adiabatic leading edge followed by a $\pi/2$ pulse [see, e.g.,
Ref.~\cite{Faust2013} and extra step I' in Fig.~\ref{fig:1}~(a)]. While such a protocol is not limited by bandwidth constraints,
it is limited by adiabaticity and the fidelity of the resonant $\pi/2$ pulse. A high-fidelity adiabatic pulse must fulfill the
condition $\Omega_0 t_\us \gg 1$. This renders high fidelity preparation of the sensing state in the presence of noise
unsustainable~\cite{nalbach2009}.

To design a fast, bandwidth-limited protocol yielding a high-fidelity state preparation [orange arrows in Fig.~\ref{fig:1}~(a)],
we start from the single-tone function [green line in Fig.~\ref{fig:1}~(d)]
\begin{equation}
	f (t)=\frac{1}{2}\left[1 + \cos\left(\frac{\pi}{t_\us} t\right)\right],
	\label{eq:InitPrepDet}
\end{equation}
and choose a $t_\us$ which respects the bandwidth limitation imposed by various experimental components including arbitrary
waveform generators (AWG), filters, amplifiers, and other passive and active circuit components in a laboratory environment. This
implies that the allowed maximal $t_\us$ in Eq.~\eqref{eq:InitPrepDet} is in general still too slow to realize a quasi
instantaneous sweep, but yet much shorter than the $t_\us$ required to fulfill the adiabatic criterion. 

In this intermediate regime, where the generated evolution is coherent, we use the recently proposed Magnus-based strategy for
control~\cite{Ribeiro2017,Roque2021} to cancel on average transitions to mode $\av_2$. This yields a function $f_\mm{mod} (t)$
that defines a modified frequency-sweep that allows one to achieve high-fidelity state preparation.

For the rest of this work we consider two different modified frequency-sweeps that we label Mod1 and Mod2.  The leading edges of
the uncorrected (green), Mod1 (red), and  Mod2 (blue) frequency-sweeps are shown in Fig.~\ref{fig:1}~(d).

To compare the performance of the different frequency-sweeps, we consider the state fidelity errors
\begin{equation}
	\begin{aligned}
		\varepsilon_{\us,j} &= 1 - \abs{\av_1^\mathsf{T} \Phi_j (t_\us) \av_1}^2, \\
		\varepsilon_{\ur,j} &= 1 - \abs{\av_1^\mathsf{T} \Phi_j^\mathsf{T} (t_\uf) \Phi_j (t_\ur) \av_1}^2, \\
	\end{aligned}
	\label{eq:FidErrPrep}
\end{equation}
associated to the preparation of the sensing and readout state, respectively.  Here $j \in \{0,1,2\}$ labels the different
frequency-sweeps:~$j=0$ is the uncorrected detuning sweep [see Eqs.~\eqref{eq:GenDetRamsey} and \eqref{eq:InitPrepDet}], and
$j=1,2$ are associated to the modified frequency-sweeps Mod1 and Mod2, respectively. The flow $\Phi_j (t)$ allows one to find the
state vector at time $t$, i.e., $\av (t) = \Phi_j (t) \av (0)$, and obeys the equation of motion (see, e.g.,
Ref.~\cite{Seitner2017})
\begin{equation}
	i \dot{\Phi}_j (t) = D_j (t) \Phi_j (t),
	\label{eq:Flow}
\end{equation}
with $D_j (t)$ given by Eq.~\eqref{eq:DynMat}. 

\begin{figure}
	\includegraphics[width=\columnwidth]{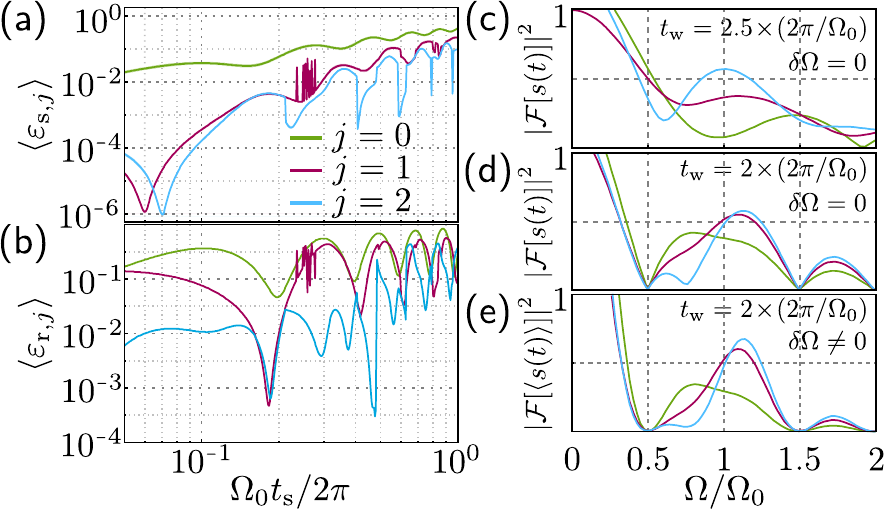}
	\caption{State preparation and trivial signal processing. (a) Comparison of the averaged sensing state error fidelity [see
		Eq.~\eqref{eq:FidErrPrep}] between the uncorrected (green), Mod1 (red), and Mod2 (blue) frequency-sweeps as a
		function of $t_\us$. (b) Same as (a) for the readout state. (c) Discrete spectral density $\abs{\mathcal{F}\langle
		s(t)\rangle}^2$ obtained from signals generated with the uncorrected ($j=0$, green), Mod1 ($j=1$, red) and Mod2
		($j=2$, blue) frequency-sweep for $t_\uw= 2.5\times (2\pi/\Omega_0)$ and $\delta \Omega =0$. (d) Same as (c) for
		$t_\uw= 2\times (2\pi/\Omega_0)$. (e) Same as (d) for $\delta \Omega  \neq 0$ [see Eqs.~\eqref{eq:NoiseAvg} and
		\eqref{eq:9}].  
	}
\label{fig:2}
\end{figure}

Figure~\ref{fig:2}~(a) and (b) display the averaged errors $\langle\varepsilon_{\us,j}\rangle$ and
$\langle\varepsilon_{\ur,j}\rangle$ [obtained by averaging $\varepsilon_{\us,j}$ and $\varepsilon_{\ur,j}$ over noise, see
Eq.~\eqref{eq:NoiseAvg}] as a function of $t_\us$, respectively. To achieve error fidelities on the order of $10^{-3}$ with the
uncorrected frequency-sweep (green trace), one would need to fulfill $\Omega_0 t_\us/(2 \pi) \ll 10^{-2}$, which would require an
AWG with a very large bandwidth. In stark contrast, the corrected detuning-sweeps (red and blue traces) allow one to achieve error
fidelities smaller than $10^{-3}$ for values of $\Omega_0 t_\us$ that are much larger than the ones required for a quasi
instantaneous sweep, which lessens the bandwidth requirements associated with such a sweep.

\textit{Frequency estimation with trivial signal processing ---}
Although our modified frequency-sweeps allow us to prepare the ideal sensing and readout states, they do not allow us to correctly
estimate unknown frequencies from short-time signals, i.e., $2 \leq (\Omega_0/2\pi) t_\uw \leq 4$.

We illustrate this in Fig.~\ref{fig:2}~(c)-(e) where we compare the modulus squared of the discrete signal Fourier transform
(spectral density), $\abs{\mathcal{F}[\langle s(t)\rangle]}^2$, for different case scenarios. We stress that the spectra were
obtained assuming that we know exactly the value of $\Omega_0$ to capture only the effects of the shortcomings associated to
Fourier transforms of short-time signals (FT1).

The modified frequency-sweeps (red and blue traces) lead to spectra where the global maximum (with the 0-frequency peak excluded)
can be more easily identified. However, the global maximum is not located at $\Omega/\Omega_0 =1$. Our results also show that
small changes in $t_\uw$ can result in different spectra with maxima located at very different frequencies, which obviates a
correct frequency estimation.

In the following, we show how an iterative procedure combining both an update on the estimate for $\Omega_0$ and different
windowing schemes~\cite{Prabhu2014} to process the measured signal solves the issues outlined above and yields a high-precision
estimate of the unknown frequency.

\textit{Iterative, adaptive frequency estimate procedure ---}
Windows, or tapers, are weighting functions designed to simplify the analysis of harmonic signals in the presence of noise and
harmonic interference. In particular, the window functions apply selective weights to reduce spectral leakage associated with
finite measurement windows~\cite{Prabhu2014,Harris1978}. 

\begin{figure}
	\includegraphics[width=\columnwidth]{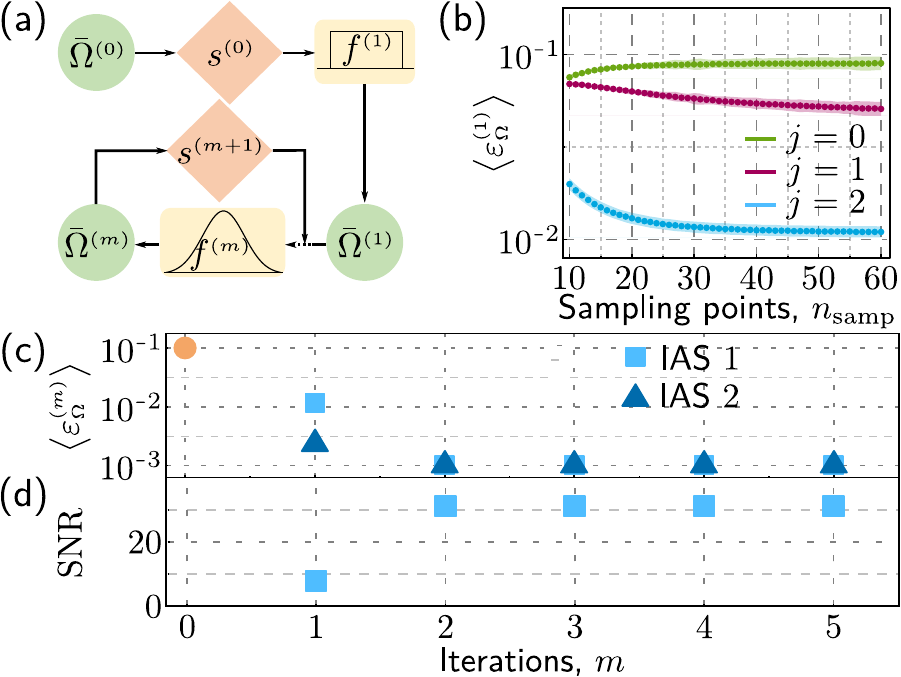}
	\caption{Iterative adaptive spectroscopy of short signals. (a) Flowchart illustrating the steps involved at each iteration. (b) Comparison of
		the relative error $\epsilon_{\Omega}^{(1)}$ of the frequency estimate as a function of sampling points
		$n_\mm{samp}$
		for the uncorrected ($j=0$, green), Mod1 ($j=1$, red), and Mod2 ($j=2$, blue) detuning-sweeps. The shaded region shows the error on the
		measurement. (c) Relative error $\epsilon_{\Omega}^{(m)}$ as a function of iterations for IAS 1 (light blue
		squares) and IAS 2 (dark blue triangles). (d) $\mm{SNR}$ [see Eq.~\eqref{eq:SNR}] as a function of iterations for IAS 1. In (c) and
		(d) the error is smaller than the symbols. Otherwise specified we use $t_\uw = 4\times (2\pi/\bar{\Omega}^{(m)})$,
		$t_\uf=0.5\times (2\pi/\bar{\Omega}^{(m)})$, and $n_\mm{samp}=30$.
	}
\label{fig:3}
\end{figure}

In this context, the so-called Blackman-Harris window~\cite{Prabhu2014,Harris1978} is notably known to effectively reduce spectral
leakage. However, as a downside, the measurement-time window must be chosen such that $(\Omega_0/2\pi) t_\uw\geq 4$ and the
amplitude of the spectral density is reduced, which can render frequency detection problematic, specially for short-time, noisy
signals. Finally, and we cannot stress this enough, while windowing reduces spectral leakage, it can never completely suppress it.
Thus, even with windowing, high-precision frequency estimation is still limited by artifacts linked to discrete Fourier transforms
of short-time signals. 

To overcome this limitation we use the iterative, adaptive sensing (IAS) protocol depicted in Fig.~\ref{fig:3}~(a). Each
interaction consists in performing Ramsey interferometry [rhombi in Fig.~\ref{fig:3}~(a)] with a frequency-sweep that takes into
account our current knowledge of the frequency estimate [circles in Fig.~\ref{fig:3}~(a)]. This way we can iteratively suppress
systematic frequency shift errors originating from spectral leakage when $t_\uw$ [step III in Fig.~\ref{fig:1}~(a)] is not an
integer multiple of the period. This is done by updating after each iteration the measurement-time window
\begin{equation}
	t_\uw^{(m+1)} = n \times \frac{2\pi}{\bar{\Omega}^{(m+1)}} \quad m \in \mathbb{N}_0,
	\label{eq:ItProc}
\end{equation}
where $\bar{\Omega}^{(m)}$ denotes the frequency estimate obtained at iteration $m$ and $n \geq 4$ is an integer that defines how
many periods are in the measurement-time window. 

We assume in Eq.~\eqref{eq:ItProc} that a prior estimate of the frequency, which we denote by $\bar{\Omega}^{(0)}$, is known. This
quantity can, e.g., be obtained with ``standard'' Ramsey interferometry or by using spectroscopic methods. 

We also update after each iteration the leading and trailing edge of the modified detuning-sweeps since the Magnus-based strategy
for control requires one to know the parameters entering the equations of motion [see Eqs.~\eqref{eq:DynMat} and \eqref{eq:Flow}].
Thus, we decrease at each iteration the error in preparing both the sensing and readout state.

A new estimate of the frequency is obtained by processing the acquired signal either with a window function tailored to either
enhance the amplitude of the Fourier spectrum or frequency estimation [rectangles in Fig.~\ref{fig:3}~(a)].  The window functions
for signal processing are chosen according to
\begin{equation}
	f^{(m)} (t) = 
	\begin{cases}
		\Theta(t) - \Theta(t-t^{(1)}_\uw) &\text{for } m =1,\\
		f_\mm{BH} (t/t^{(m)}_\uw -1/2) &\text{for } m\geq 1,
	\end{cases}
	\label{eq:windows}
\end{equation}
where $\Theta(t)$ denotes the Heaviside function and $f_\mm{BH} (t)$ is the Blackman-Harris window. We deliberatively choose not
to process the signal after the first iteration (we use a rectangular window)
\footnote{From the point of view of signal processing applying a rectangular window is the same as applying no window.}
in order to generate a spectral density with larger amplitudes, from which maxima are easier to extract.

The other source of systematic errors come from dealing with a signal which is constructed from a finite number of sampling points
(measurements) $n_\mm{samp}$ (FT2). A small $n_\mm{samp}$ will lead to scalloping, i.e., the $\floor{n_\mm{samp}/2}$-point
discrete Fourier transform does not resolve the real maxima of the spectrum~\cite{Prabhu2014}. This is, however, easily fixable by
using zero-padding, as described in signal processing textbooks, e.g., in Ref.~\cite{Prabhu2014}.  Zero-padding consists in
extending the signal with $n_\mm{pad}$ zeros yielding a $\floor{(n_\mm{samp} + n_\mm{pad})/2}$-point discrete Fourier transform.
To further reduce the effects of scalloping we use interpolation of the padded discrete spectrum~\cite{gasior2004} (see also
Supplemental Material).

Figure~\ref{fig:3}~(b) shows the relative error of the frequency estimate
$\epsilon^{(1)}_{\Omega}=\log_{10}(\abs{1-\bar{\Omega}_{(1)}/\Omega_0})$ as a function of $n_\mm{samp}$ for $n=4$ for the first
iteration [$m=1$ in Eqs.~\eqref{eq:ItProc} and \eqref{eq:windows}] of our iterative sensing scheme. Here $n_\mm{pad}$ is chosen
such that $n_\mm{samp} + n_\mm{pad} = 1000$. Independently of the frequency-sweep used, doubling $n_\mm{samp}$ only leads to a
small variation of the relative error. This allows us to identify $\ceil{n_\mm{samp}/n} = 8 > 2$ as a good compromise between the
error and experimental cost, i.e., the number of measurements. The results show the advantage of using the modified
frequency-sweeps Mod1 (red trace) and Mod2 (blue trace) over the uncorrected one (green trace); the smaller the error in preparing
both the sensing and readout state the smaller the relative error of the first frequency estimate for a signal of identical
duration. 

In Fig.~\ref{fig:3}~(c) we plot the relative error $\epsilon^{(m)}_{\Omega}$ after each iteration of our adaptive scheme (IAS 1,
light blue squares). We also included the error on our prior estimate $\bar{\Omega}^{(0)}$ for reference (orange circle). The
iterative procedure converges to a value of the frequency estimate whose error is smaller than the initial estimate. Convergence
indicates that we reached the spectral resolution allowed by our measurement time-window after just a few iterations.

To show that the choice of window for the first iteration has no influence on the results, we also plotted in Fig.~\ref{fig:3}~(c)
the relative error obtained by using at every step the Blackman-Harris window for signal processing (IAS 2, dark blue triangles).

Finally, we plot in Fig.~\ref{fig:3}~(d) the signal-to-noise (SNR) ratio defined as
\begin{equation}
	\mm{SNR}=\frac{\sqrt{N} \mathcal{F}[ \avg{s^{(n)} (t) f^{(n)}(t)}]} {\left[\sum\limits_{j=1}^{N} \left( \mathcal{F}[
	s_j^{(n)} (t) f^{(n)}(t)]- \mathcal{F}[ \avg{s^{(n)} (t) f^{(n)}(t)}]\right)^{2}\right]^{1/2}},
	\label{eq:SNR} 
\end{equation}
which is a measure of the confidence level on the frequency estimate. More precisely, in this context, the $\mm{SNR}$ quantifies
the degree of confidence we have in identifying the global maximum of the spectral density (0-frequency component excluded). Our
results show that as the relative error on the frequency estimate decreases, we become more and more confident in identifying the
frequency associated to the global maximum of the spectral density in spite of having a noisy signal. 

\textit{Conclusion ---}
We have developed an iterative, adaptive sensing protocol based on enhanced Ramsey interferometry of two-level systems. Our scheme
allows one to get precise estimates of an unknown frequency by considering short, finite-time signals under realistic assumptions
of experimental bandwidth limitations. Specifically, our scheme avoids shortcomings both related to dealing with decaying signals
and experimental constraints related to the sampling and could be implemented, e.g., in coupled mechanical
oscillators~\cite{Faust2012b,Okamoto2013a,Seitner2017b,Braakman2018}, optomechanical systems~\cite{Teufel2011,Ranfagni2021},
hybrid optomechanical systems~\cite{Yeo2014}, coupled optical modes~\cite{Thon2009}, and qubits~\cite{oliver2005,petta2010} under
the influence of classical noise, just to name a few.  

The main ingredients of our method are the use of the Magnus-based strategy for control to find frequency-sweeps that allow one to
prepare with high fidelity both the sensing and readout state and an iterative procedure built to mitigate systematic errors when
using Fourier transforms to extract frequency components. We stress that independently of how the sensing and readout state are
prepare, our iterative, adaptive sensing protocol can always be applied to enhance frequency estimates.

%Code used to calculate the Magnus-based correction is available at~\cite{CodeLink}.

%\bibliography{iterative_ramsey}

%

\begin{appendix}
\clearpage
\thispagestyle{empty}
\onecolumngrid
%\clearpage
\begin{center}
\textbf{\large Supplemental Material:\\
Iterative Adaptive Spectroscopy of Short Signals}
\end{center}

\setcounter{secnumdepth}{2}
\setcounter{figure}{0}
\renewcommand{\thefigure}{S\arabic{figure}}

\section{Using $\av_1$ as sensing state}

In this section we briefly show the advantages of using the sensing state $\av_\us = \av_1$. 

The advantage is that $\av_1$ maximizes the visibility of the Ramsey signal. This can be readily verified by assuming that one can
prepare with unit fidelity both the sensing and readout state. In this case the Ramsey signal is given by 
\begin{equation}
	s (t) = \abs{\av_1^\mathsf{T}  \exp(-i \Omega_0 t \sigma_x) \av_1}^2 = \cos^2\left[ (\Omega_0/2) t  \right],
	\label{eq:TrivialRamsey}
\end{equation}
which is a function oscillating between $0$ and $1$, and thus with unit visibility. 

To understand why this property is important in the context of frequency estimation from short signals, let us consider the
situation where we prepare any arbitrary state at $t=t_\us$. We model this situation by describing the evolution generated by
$D(t)$ [see Eq.~\eqref{eq:DynMat}] between $t=0$ and $t=t_\us$ as a rotation $R_{\nvs} (\theta)$ of angle $\theta$ around an axis
$\nv = (\sin \alpha \cos \beta, \sin \alpha \sin \beta, \cos \beta)^\mathsf{T}$. We have 
\begin{equation}
	R_{\nvs} (\theta) = \cos(\theta/2) \mathbbm{1} - i \sin(\theta/2) \nv \cdot \sigmav,
	\label{eq:Gen3DRot}
\end{equation}
with $\sigmav = (\sigma_x, \sigma_y, \sigma_z)^\mathsf{T}$ the vector of Pauli matrices. Similarly, we assume that the preparation
of the readout state is described by the rotation $R_{\nvs} (-\theta)$. Within this framework, the Ramsey signal is given by 
\begin{equation}
	\begin{aligned}
		s_\theta (t) &= \abs{\av_1^\mathsf{T}  R_{\nvs} (-\theta) \exp(-i \Omega_0 t \sigma_z) R_{\nvs} (\theta) \av_1}^2 \\
		&= 1 + \left[ \sin^2 \alpha \left(2 \cos \alpha \cos \beta \sin^2(\theta/2) + \sin \beta \sin \theta \right)^2 -1
		\right] \sin^2\left[ (\Omega_0/2) t \right] \\
		&= 1 + v(\alpha, \beta, \theta)  \sin^2\left[ (\Omega_0/2) t \right]
	\end{aligned}
	\label{eq:ToyModelRamseySig}
\end{equation}
For $\alpha = 0$ with $\beta$ and $\theta$ arbitrary, we have $s_\theta (t) = s (t)$ since this corresponds to a rotation around
the $z$-axis which only imprints a global phase to mode $\av_1$. Another situation that leads to $s_\theta (t) = s (t)$ is by
setting $\alpha = \pi/2$, $\beta=0$ and $\theta$ arbitrary, which corresponds to a rotation around the $x$-axis. This does also
not affect the Ramsey signal since the Ramsey signal is obtained by letting the state vector precess freely around the $x$-axis. 

For any other rotation, the Ramsey signal can be viewed as the sum of two signals:~A constant signal $s_1 (t) =1$  and an
oscillating signal $s_2 (t) = v(\alpha, \beta, \theta)  \sin^2\left[ (\Omega_0/2) t \right]$  with visibility $0 \leq
\abs{v(\alpha, \beta, \theta)}  < 1$.  Thus, we can write the Fourier spectrum as 
\begin{equation}
	\abs{\mathcal{F}[s_\theta (t)]}^2 = \abs{\mathcal{F}[s_1 (t)]}^2 + \abs{\mathcal{F}[s_2 (t)]}^2 + 2 \mm{Re}\left[
	\mathcal{F}[s_1 (t)] \mathcal{F}[s_1 (t)]^\ast \right].
	\label{eq:FourierSumSigs}
\end{equation}

Equation~\eqref{eq:FourierSumSigs} shows that one can interpret the spectrum of a finite $s_\theta (t)$ as an interferometric
pattern. As a result, $\abs{\mathcal{F}[s_\theta (t)]}^2$ does not necessarily have a maximum located at $\omega = \Omega_0$. This
is yet another type of systematic error in the sensing result which cannot be eliminated unless one knows exactly what the values
of $\alpha$, $\beta$, and $\theta$ are. 

Finally, we note that in the infinite measurement-time window limit, i.e., $t_\uw \to \infty$, this issues vanishes since the
discrete Fourier transform of $s_1 (t)$ would reduce to the Kronecker delta function. Consequently, in the limit defined by
$\Omega_0 t_\uw \gg 1$, the induced systematic error is negligible. 

\section{Magnus-based strategy for control}

In this section, we detail how we obtained the modified protocols, Mod1 and Mod2, using the Magnus-based strategy for
control~\cite{Ribeiro2017,Roque2021}. In particular, we show how we find the modification for the leading and trailing edge of the
frequency-sweep. 

The first step consists in finding a partition of the dynamical matrix $D(t) = D_0 (t) + V(t)$ (see Eq.~\eqref{eq:DynMat} of the
main text), where $D_0 (t)$ generates the desired dynamics and $V(t)$ describes the spurious coupling disrupting the desired
dynamics.  For the problem at hand, and neglecting noise, we have 
\begin{equation}
	D_0 (t) = \frac{1}{2} \Delta (t) \sigma_z,
	\label{eq:D0}
\end{equation}
and 
\begin{equation}
	V(t) = \frac{1}{2} \Omega_0 \sigma_x.
	\label{eq:Vbad}
\end{equation}

The second step consists in introducing a control $W (t)$ which cancels on \emph{average} the spurious effects generated by
$V(t)$. Formally, this leads to a modified dynamical matrix $D_\mm{mod} (t) = D(t) + W (t)$, which generates a flow $\Phi_\mm{mod}
(t)$. The control $W (t)$ must be chosen such that 
\begin{equation}
	\Phi_\mm{mod} (t_\uf) = \Phi_0 (t_\uf),
	\label{eq:Wcond}
\end{equation}
where $\Phi_0 (t)$ is the flow generated by $D_0 (t)$ [see Eq.~\eqref{eq:D0}].

Following the procedure detailed in~\cite{Roque2021}, we consider $W(t) = \Delta_\mm{corr} (t) \sigma_z$, which is compatible with
the constraints of the problem; we can only control in time the field coupling to $\sigma_z$. Taking advantage of the mirror
symmetry of the frequency-sweep around $t=t_\ur /2$, we can express $\Delta_\mm{corr} (t)$ as 
\begin{equation}
	\Delta_\mm{corr}(t) =
	\begin{cases}
		\Delta_\mm{even} (t) + \Delta_\mm{odd} (t)  &\text{for } 0 \leq t \leq t_\us, \\
		0 &\text{for } t_\uf < t < t_\ur, \\
		\Delta_\mm{even} (t) - \Delta_\mm{odd} (t)  &\text{for } t_\uf \leq t \leq t_\ur,
	\end{cases}
	\label{eq:DetCorr}
\end{equation}
where $\Delta_\mm{even} (t)$ and $\Delta_\mm{odd} (t)$ are parametrized as finite Fourier series 
\begin{equation}
	\begin{aligned}
		\Delta_\mm{even} (t) &=\sum_{k=1}^{k_\mm{max}}c_{k}\left[ 1-\cos{\left(2\pi k\frac{t}{t_\uf}\right)}\right],\\
		\Delta_\mm{odd} (t) &=\sum_{l=1}^{l_\mm{max}}d_l \sin{\left(2\pi l\frac{t}{t_\uf}\right)}.
	\end{aligned}
	\label{eq:EvenOddFourier}
\end{equation}
Here $c_k$ and $d_l$ are the free Fourier coefficients one must find in order to fulfill Eq.~\eqref{eq:Wcond}. The number of free
coefficients is set by choosing appropriate values for $k_\mm{max}$ and $l_\mm{max}$, e.g., one might want to constraint
bandwidth.  We stress that we parametrized Eq.~\eqref{eq:DetCorr} such that the coefficients $c_k$ and $d_l$ are the same for the
leading and trailing edge of $\Delta_\mm{corr}(t)$. 

Equations for $c_k$ and $d_l$ are found by transforming $D_\mm{mod} (t)$ to the interaction picture defined by $\Phi_0 (t)$, i.e.,
$D(t) \to D_\mm{mod,I} (t) = \Phi_0^\dag (t) D_\mm{mod} (t) \Phi_0 (t) - i \Phi_0^\dag (t) \dot{\Phi}_0 (t)$. We find
$D_\mm{mod,I} (t) = V_\uI (t) + W_\uI (t)$, where 
\begin{equation}
	V_\uI (t) = \frac{\Omega_0}{2}\left[\cos\left( \int_0^t \di{t_1} \Delta(t_1) \right)\sigma_x - \sin\left( \int_0^t
	\di{t_1} \Delta(t_1) \right)\sigma_y  \right]
	\label{eq:VI}
\end{equation}
and
\begin{equation}
	W_\uI (t) = W (t).
	\label{eq:WI}
\end{equation}

By comparing $W_\uI (t)$ with $V_\uI (t)$, we notice that the control dynamical matrix $W(t)$ is singular~\cite{Roque2021}. Thus,
we follow the strategy outlined in section ``Singular or ill-conditioned correction Hamiltonians'' of Ref.~\cite{Roque2021} to
obtain a set of nonlinear equations for the coefficients $c_k$ and $d_l$. 

Here, we look for the coefficients $c_k$ and $d_l$ which fulfilled the coupled equations 
\begin{equation}
	\begin{aligned}
		\frac{1}{2}\mm{Tr}\left[\sum_{k=1}^4 \Omega_k^{(1)} (t_\uf) \sigma_x \right] &= 0,\\
		\frac{1}{2}\mm{Tr}\left[\sum_{k=1}^4 \Omega_k^{(1)} (t_\uf) \sigma_y \right] &= 0,
	\end{aligned}
	\label{eq:4thOrdCoeffs}
\end{equation}
where $\Omega_k^{(1)} (t_\uf)$ are the elements of the Magnus series generated by $D_\mm{mod,I} (t)$. Since we only want to
prevent (coherent) transitions on average from mode 1 to mode 2 and vice versa, we only look for the coefficients $c_k$ and $d_l$
that cancel the off-diagonal elements of the Magnus expansion up to fourth order.

Since the non-linear system of equations in Eq.~\eqref{eq:4thOrdCoeffs} has in general more than one solution, one can choose the
solution that minimizes the norm of the vector of free parameters, i.e., the function $g = \sum_{k,l} (c_k^2 + d_l^2)$. In
particular, it is numerically more efficient to directly minimize $g$ under the constraints defined by
Eq.~\eqref{eq:4thOrdCoeffs}. 
%(code available at \cite{CodeLink}).

\section{Bandwidth limitation of the linear ramp pulse}

\begin{figure}[h]
	\begin{center}
		\includegraphics[width=0.4\columnwidth]{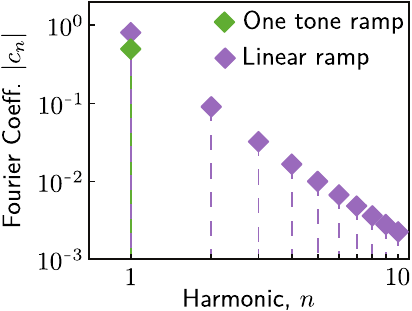}
	\end{center}
	\caption{Plot of the Fourier amplitude $\abs{c_\mm{n}}$ [see Eq.~\eqref{eq:S1}] as a function of the harmonic $n$, for one
	tone (green) and linear ramp $s_\mm{triangle}(t)$ (purple).}
	\label{fig:S1}
\end{figure}

In this section we briefly discuss the bandwidth requirements to use a linear sweep function $f(t)$ (see
Eq.~\eqref{eq:GenDetRamsey} of the main text). Linear ramps are obtained by approximating a triangular wave with a period equal to
$T$. The linear ramp is then obtained by considering the time-interval $t \in [0,t_\us=T/4]$. The Fourier series of the triangle
wave is given by
\begin{equation}
	\begin{aligned}
		s_\mm{triangle} (t) &=\sum_{n=1}^{\infty}c_n\sin\left(2\pi (2n-1) \frac{t}{T}\right), \\
		c_n &=-\frac{8}{\pi^{2}} \frac{(-1)^n}{(2n-1)^2}\,. 
	\end{aligned}
\label{eq:S1}
\end{equation}
This indicates that one needs to use an infinite number of harmonics to faithfully reproduce a linear ramp (see also
Fig.~\ref{fig:S1}).

\section{State preparation errors}

In this section we discuss how noise and low fidelity state preparation affects the sensing protocol.

\begin{figure}[h!]
	\begin{center}
		\includegraphics[width=0.8\columnwidth]{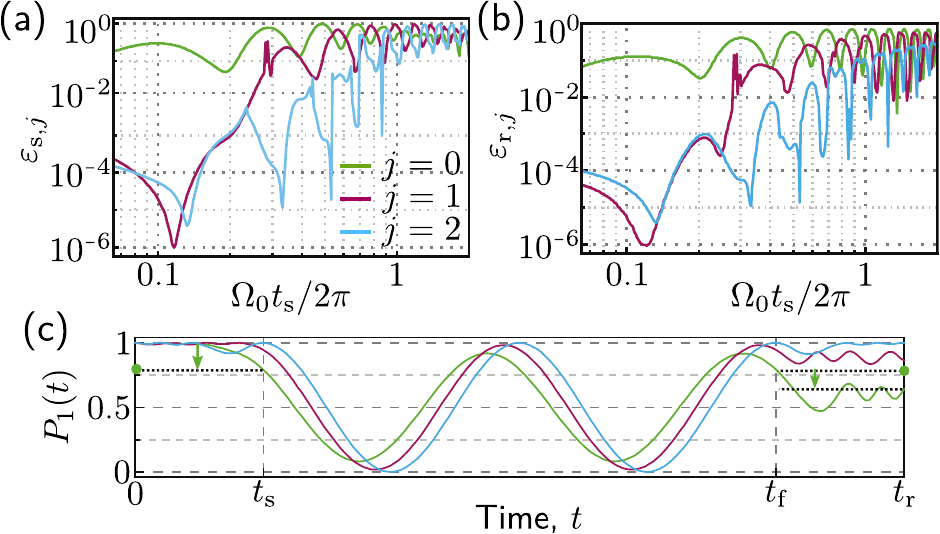}
	\end{center}
	\caption{(a) Comparison of the sensing state error fidelity for $\delta \Omega =0$ between the uncorrected (green) Mod1
	(red), and Mod2 (blue) frequency-sweeps as a function of $t_\us$. (b) Same as (a) for the readout state. (c) Evolution of
	the probability $P_1 (t)$ of measuring mode $\av_1$ as a function of time for $\Omega_0 t_\us/2\pi = 0.5$. The green dots and
	arrows are a visual indicator to show how much the ideal coherent evolution is corrupted by coherent errors when using the
	uncorrected pulse in both sensing and readout state.}
	\label{fig:S2}
\end{figure}

\subsection{Noise-induced errors}

In the main text, we showed the noise averaged sensing and readout state fidelity error (see Fig.~\ref{fig:2}~(a) and (b) of the
main text). Here, to give a sense on how noise hinders =the preparation of both states, we present in Fig.~\ref{fig:S2}~(a) and
(b) $\varepsilon_{\us, j}$ and $\varepsilon_{\ur, j}$ (see Eq.~\eqref{eq:FidErrPrep} of the main text) obtained in the absence of
noise ($\delta \Omega =0$ in Eq.~\eqref{eq:DynMat} of the main text).

Let us first consider the sensing state fidelity error for the uncorrected protocol [green trace in Fig.~\ref{fig:2}~(a) of the
main text and Fig.~\ref{fig:S2}~(a)]. In this case, the presence of noise favors the state preparation by preventing the
generation of the ideal coherent evolution and thus partially suppressing coherent transitions to mode $\av_2$. On the other hand,
for the modified protocols [red and blue traces in Fig.~\ref{fig:2}~(a) of the main text and and Fig.~\ref{fig:S2}~(a)] that rely
on coherent evolution to average out the effects of the spurious interaction [see Eqs.~\eqref{eq:Vbad} and \eqref{eq:VI}] noise
reduces the state preparation fidelity. However, as the results of the main text show, the fidelity errors obtained with the
modified protocols in the presence of noise are still orders of magnitude smaller than the uncorrected one. This fact can be
attributed to having protocols that are designed to be shorter than the decoherence time set by the noise.

For the readout state preparation, one would expect the same observations as above. This is, however, not the case. Comparison of
Fig.~\ref{fig:2}~(b) and Fig.~\ref{fig:S2}~(b) reveals that noise hinders the readout state preparation to a much greater extent
than it does for the sensing state preparation. This difference originates from the dependence of the state fidelity error on the
initial state, which for preparing the sensing state is simply $\av_1$ and for preparing the readout state is $\av (t_\uf) =
\Phi_j (t_\uf) \av_1$. The latter is a coherent superposition state and is therefore more susceptible to noise-induced
decoherence.

\subsection{Coherent errors}

To illustrate how low fidelity state preparation affects the sensing protocol, we plot in Fig.~\ref{fig:S2}~(c) the probability
$P_1 (t_\uw, t)$ of measuring mode $\av_1$ as a function of time for a fixed measurement-time window $t_\uw$. We note that the
Ramsey signal $s (t_\uw)$ is constructed from the values of $P_1 (t_\uw, t_\ur)$ when $t_\uw$ is varied, i.e., $s(t_\uw) = P_1
(t_\uw, t_\ur)$. We have 
\begin{equation}
	P_{1,j} (t)=\abs{a_{1}^{T} \Phi_j (t) a_{1}}^{2},
\label{eq:S4}
\end{equation}
where $j\in\{0,1,2\}$ labels, as in the main text, which detuning sweep is used to obtain $\Phi_j (t)$. We recall that $j=0$
labels the uncorrected detuning-sweep, while $j=1,2$ labels the detuning-sweep coined Mod1 and Mod2, respectively. 

The uncorrected detuning-sweep [green trace in Fig.~\ref{fig:S2}~(c)] shows how coherent errors propagate and lead to the
``wrong'' Ramsey signal. Using Mod1 (red trace) or Mod2 (blue trace) which allow for high fidelity state preparation of both the
sensing and readout state, coherent errors are reduced and a more faithful Ramsey signal can be constructed. 

\section{Spectral leakage and scalloping losses}

\begin{figure}[h!]
	\begin{center}
		\includegraphics[width=0.9\columnwidth]{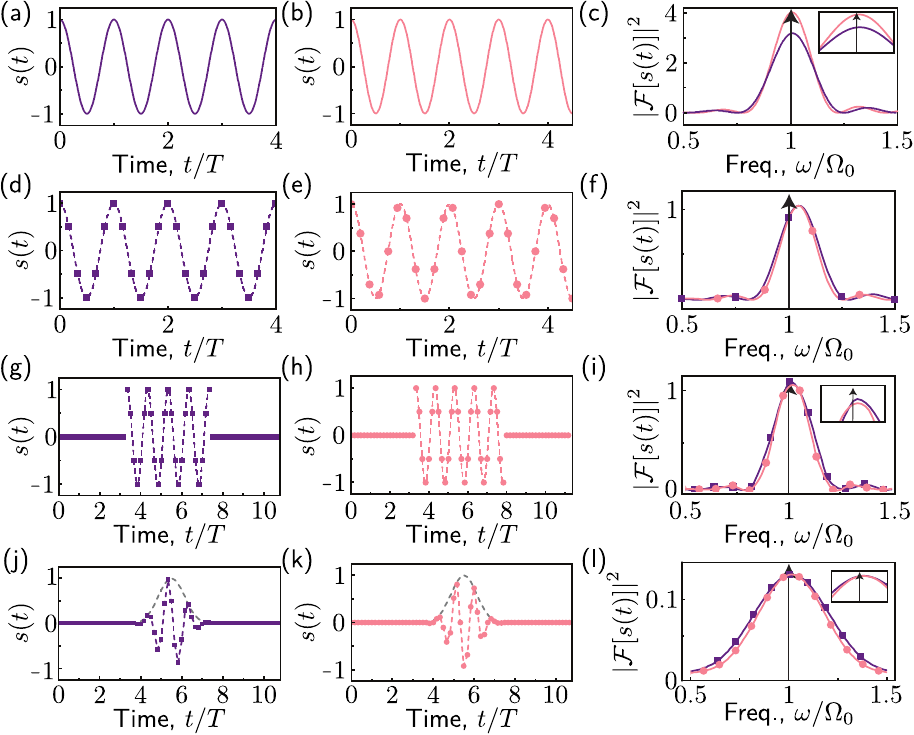}
	\end{center}
	\caption{Time and frequency domain representation of $s_\mm{1}(t)$ for a time interval of $4T$ (purple) and $4.5T$ (pink).
	(a-b) Continuous time measurement with finite time windows and their corresponding continuous Fourier transforms in (c).
	(d-e) Discrete sampling of a continuous signal for both time intervals. The associated discrete Fourier transforms (dots) and
	discrete-time Fourier transforms (solid lines) are shown in (f). (g-h) 0-padding of the sampled signal and corresponding
	interpolated discrete Fourier transforms in (i). Post-processing of the sampled signal with a  Blackmann-Harris window and
	0-padding (j-k). The resulting discrete Fourier transform is less susceptible to spectral leakage (l).}
	\label{fig:S3}
\end{figure}

To visualize the shortcomings associated with Fourier transforms of short signals, it is useful to consider a simple sinusoidal
signal 
\begin{equation}
	s_1 (t) = \cos\left( \Omega_0 t \right),
	\label{eq:CosSignal}
\end{equation}
over a finite interval of time. We consider two different intervals of time defined as $t_{1,\mm{max}} = 4 T$ and $t_{2,\mm{max}}
= 4.5 T$, where $T = 2\pi/\Omega_0$ [see Fig.~\ref{fig:S3} (a)-(b)].

Let us first consider a situation where one could continuously measure the signal. Since having a finite-time signal is equivalent
to the pointwise multiplication of an infinite signal with a rectangular window, the Fourier transform produces a a spectrum whose
value at $\omega=\omega_0$ is the sum of all the spectral contributions of the signal weighted by the spectrum of the window
centered at $\omega_0$ (convolution theorem). As a result, even a simple, single frequency spectrum appears with multiple
frequency components [see Fig.~\ref{fig:S3} (c)].

Experimentally, however, it is not always possible to measure a continuous signal, but the signal can be sampled at a certain rate
[Fig.~\ref{fig:S3} (d) and (e)]. In this situation, one uses the the discrete Fourier transform to extract information about
frequency components. This results in a spectrum with a finite number of points [dotted points in Fig.~\ref{fig:S3} (f)], from
which frequency estimation is limited due to having a finite frequency resolution set by $\delta f=1/t_\mm{max}$. This is known as
scalloping loss. 

The continuous lines in Fig.~\ref{fig:S3} (f) show the discrete-time Fourier transform associated to the sampled signals in (d-e).
The discrete Fourier transforms [dots in Fig.~\ref{fig:S3} (f)] samples the continuous spectrum generated by the discrete-time
Fourier transform with a frequency interval $\delta f$.

To minimize scalloping losses, the common practice requires one to use 0-padding [Fig.~\ref{fig:S3} (g) - (h)], which effectively
reduces $\delta f$. Additionally, 0-padding forces the signal to appear aperiodic, which mitigates spectral leakage when the time
interval is not an integer multiple of the period  [see Fig.~\ref{fig:S3} (i)]. However, the accuracy of the frequency estimate is
still limited by the use of a rectangular window (or no-window), which as we discussed above, also induces spectral leakage [inset
in Fig.~\ref{fig:S3} (i)].

Using specific window functions, e.g., the  Blackmann-Harris window [Fig. \ref{fig:S3}~(j)-(k)]~\cite{Harris1978}, one further
mitigates spectral leakages. This leads to a better estimation of the signal frequency as seen in Fig.~\ref{fig:S3} (l). As
discussed in the main text, using interpolation of the Fourier spectrum further enhances the accuracy of the frequency estimate.

\section{Iterative procedure for an idealized Ramsey signal}

In this section, we show that the iterative, adaptive sensing protocol we developed in the main text is less efficient when using
rectangular windows only.

To show this, we use a toy model which assumes an ideal Ramsey signal, $s(t)=\cos^{2}[(\Omega_{0}/2)t]$ and consider three
different IAS protocols. We consider the protocols IAS 1 and 2 discussed in the main text along with the the protocol IAS 3, which
consists in using only rectangular windows. For convenience, they are defined again below:

\begin{description}[font=$\bullet$]
\item[~IAS 1]  $f^{(m)}=\Theta(t) - \Theta(t-t^{(1)}_\uw)$ for $m=1$ and $f^{(m)}=f_{BH}(t/t_\uw-1/2)$
for $m\geq1$,

\item[~IAS 2] $f^{(m)}=f_{BH}(t/t^{(m)}_\uw-1/2)$ $\forall m$,
 
\item[~IAS 3] $f^{(m)}=\Theta(t) - \Theta(t-t^{(1)}_\uw)$ $\forall m$.

\end{description}

We assume the starting estimate to be $\bar{\Omega}^{(0)} = 1.1\Omega_{0}$ and, as in the main text, after each iteration we
update the measurement time-window $t_\mm{w}$ based on the new estimate. 

\begin{figure}[h!]
	\begin{center}
		\includegraphics[width=0.8\columnwidth]{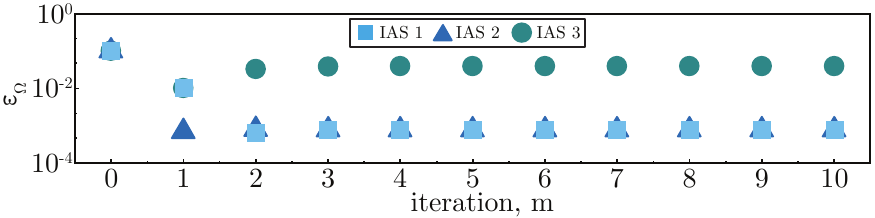}
	\end{center}
	\caption{Relative error $\epsilon_{\Omega}^{(m)}$ as a function of iteration number, $m$ for three different protocols.}
	\label{fig:S4}
\end{figure}

In Fig.~\ref{fig:S4} we show the relative error of the frequency estimate (see main text) using the previously defined IAS
protocols. Using IAS 3 results in a less accurate estimation due to using rectangular windows, which lead to spectral leakage. The
results generated by IAS 1 and IAS 2 lead to the same relative error for $m>1$, in agreement with the results shown in the main
text. 

As explained in the main text, we find IAS 1 to be the protocol of choice, since it allows one to make the first frequency
estimate from a spectrum whose main frequency component has a larger Fourier amplitude.

\end{appendix}

\end{document}